\documentclass[12pt,preprint]{aastex}

\newcommand{\eg}{e.g.,}

\newcommand{\etal}{{et al.}}
\newcommand{\kms}{{km~s$^{-1}$}}
\newcommand{\pmpc}{{Mpc$^{-1}$}}
\newcommand{\sersic}{S\'{e}rsic}
\newcommand{\jzo}{SDSSJ0114}
\newcommand{\jtt}{SDSSJ2329}
\newcommand{\msun}{$M_{\odot}$}

\begin{document}

\title{Two Moderate-Redshift Analogs to Compact Massive Early-Type Galaxies at High Redshifts\altaffilmark{1}}
\author{Alan Stockton, Hsin-Yi Shih, and Kirsten Larson}
\affil{Institute for Astronomy, University of Hawaii, Honolulu, HI 96822; stockton@ifa.hawaii.edu, hsshih@ifa.hawaii.edu, klarson@ifa.hawaii.edu}

\altaffiltext{1}{Some of the data presented herein were obtained at the W.M. Keck Observatory, which is operated as a scientific partnership among the California Institute of Technology, the University of California and the National Aeronautics and Space Administration. The Observatory was made possible by the generous financial support of the W.M. Keck Foundation.
}

\begin{abstract}
From a search of a portion of the sky covered by the SDSS and UKIDSS databases, we have located 2 galaxies at $z\sim0.5$ that have properties similar to those of the luminous passive compact galaxies found at $z\sim2.5$. From Keck moderate-resolution spectroscopy and laser-guided adaptive-optics imaging of these galaxies, we can begin to put together a more detailed picture of what their high-redshift counterparts might be like. Spectral-synthesis models that fit the $u$ to $K$ photometry also seem to give good fits to the spectral features. From these models, we estimate masses in the range of 3--$4\times10^{11} M_{\odot}$ for both galaxies.  Under the assumption that these are spheroidal galaxies, our velocity dispersions give estimated masses about a factor of 3 smaller. However, our high-resolution imaging data indicate that these galaxies are not normal spheroids, and the interpretation of the kinematic data depends critically on the actual morphologies and the nature of the stellar orbits. While recent suggestions that the population of high-redshift compact galaxies is present locally as the inner regions of local massive elliptical galaxies are quite plausible, the peak mass surface densities of the two galaxies we discuss here appear to be up to a factor of 10 higher than those of the highest density local ellipticals, assuming that our photometric masses are roughly correct. It thus seems possible that some dynamical ``puffing-up'' of the high-redshift galaxies might still be required in this scenario.
\end{abstract}


\section{Introduction}
Luminous passive galaxies at $z>2$ generally are very compact, with $R_e\lesssim2$ kpc \citep[\eg][]{dok08,dam09}. Many of these have nearly exponential profiles and appear to be disk-like \citep{sto04,sto07}, but at least some of the most compact ones, typically with $R_e<1$ kpc, seem to have surface-brightness profiles closer to an $r^{1/4}$ law \citep[\eg][]{dok08}. The most extreme example that we know of is a galaxy with an apparent mass of $\sim3\times10^{11}$ \msun\ and an $R_e=440$ pc at $z=2.48$ (Stockton \etal, in preparation). Galaxies such as these are absent, or at least extremely rare, in the present-day universe, and much of the recent discussion has centered on either the reliability of the size estimates at high redshifts or the evolutionary path such galaxies might have taken and how their stars have been incorporated into galaxies in the local universe \citep[\eg][]{dav08,cim08,fan08,bez09,hop09}. 

It is extremely difficult to study these luminous, compact galaxies at $z\sim2.5$ beyond obtaining their photometric SEDs, which confirms their approximate redshifts, and high-resolution images (HST or adaptive optics), which can indicate their gross morphologies. Photometric masses can be estimated from model fits to the SEDs, but these are uncertain because of partial degeneracies among age, metallicity, and reddening, as well as basic uncertainties in the models \citep[\eg][]{muz09}. Recently, \citet{dok09} have claimed a very high velocity dispersion of 510 \kms\ for a compact galaxy at $z=2.2$ from 29 hours of observation with GNIRS on Gemini South, but this observation only emphasizes the heroic efforts that are needed with current facilities to obtain such results, which, even so, are not immune to worries that unrecognized systematic effects may jeopardize any conclusions reached. It would be extremely useful to find similar galaxies at lower redshifts, where they could be studied in greater detail.

There have been some recent efforts to identify such galaxies. \citet{tay09} searched the Sloan Digital Sky Survey (SDSS) database for early-type compact galaxies in the redshift range $0.066<z<0.12$, using both SDSS spectra and photometric redshifts. They found a number of galaxies with $1.1<R_e<1.5$, but these all had indicated masses of $\sim5\times10^{10}$ \msun, so they are not really comparable to the massive compact galaxies found at high redshifts. \citeauthor{tay09} conclude that such galaxies must be extremely rare at the present epoch and that their size evolution cannot be a result of a stochastic mechanism such as major mergers.  On the other hand, \citet{val09} claim to find substantial numbers of massive, old, compact galaxies in nearby X-ray-selected clusters, including a few with $M>10^{11}$ \msun\ and $R_e < 1.5$ kpc.  \citeauthor{val09} use  a circularized value for $R_e$, rather than the more common major axis value (which we use throughout this {\it Letter}), but it is not clear how much this affects their conclusions.

We present here some initial results from a search at $z\sim0.5$ of a portion of the area common to the SDSS and the UKIRT Infrared Sky Survey (UKIDSS) surveys, describing in some detail results we have obtained for two of the galaxies that have properties within the range of the passive compact galaxies found at high redshifts. Throughout, we assume a flat cosmology with $H_0 = 71$ \kms\ \pmpc\ and $\Omega_m = 0.27$.

\section{Object Selection and Observations}
We selected objects from the area common to the SDSS DR7 and the UKIDSS DR5plus for RAs between $21^{\rm h}$ and $2^{\rm h}$, a total area of $\sim200$ deg$^2$.
In this region, we searched the combined UKIDSS/SDSS database for objects (1) that had SDSS $m_{i} {\rm (model)} - m_i {\rm PSF} < 0.3$  (this criterion ensures that the best-fitting model for the object is close to a PSF profile), (2) that had colors expected for old stellar populations at a given redshift, and (3) that had magnitudes that would place them at or above $\sim2 L^*$ at that redshift.  We made separate searches at redshifts ranging from 0.40 to 0.60, stepping in intervals of 0.05 in $z$. To pass initial selection, an object had to match an old-stellar population SED in 7 colors constructed from adjacent filters in the SDSS and UKIDSS surveys (no colors crossing the surveys were used), where the fitting range was $\pm3$ times the typical uncertainty, estimated separately for each color. The only serious contaminating objects we have identified are certain varieties of carbon stars, which tend to mimic old-stellar-population colors at $z\sim0.4$ surprisingly well over the 9 bands 
we have from the UKIDSS/SDSS databases. We made a final prioritization of objects for further observation by evaluating the galaxies' profiles in the SDSS $i$-band images with {\sc galfit} \citep{pen02} and eliminating those with apparent $R_e > 1.5$ pixels. The final sample comprised 14 objects. The two galaxies we discuss here, SDSS J232949.60+151106.3 (hereinafter \jtt) and SDSS J011436.33$-$011438.1 (hereinafter \jzo), are both classified as ``stars'' in the SDSS database.

We carried out spectroscopy of 11 of the 14 candidate galaxies on 2009 Aug 22 UT and 2009 Sep 16 UT with the Low-Resolution Imaging Spectrograph (LRIS; \citealt{oke95}) on the Keck I telescope.  The spectra were obtained with the 600/5000 grating and a 1\arcsec-wide slit. The detector consisted of 2 LBL $2048\times4096$ CCDs, binned $2\times2$. This configuration gives a resolution of $\sim5$ \AA. Most of the objects were, in fact, galaxies at redshifts ranging from 0.4 to 0.8. Three were carbon stars, and 1 was a peculiar QSO.

We also obtained $H$-band images of 6 of these galaxies on the night of 2009 Sep 15 UT with the Keck II laser-guide-star adaptive-optics system (LGSAO; \citealt{wiz06}) and the NIRC2 camera. Total exposures for each of the two objects discussed here were 18 min at the 40 mas pixel scale. Before registering and co-adding the individual exposures, we corrected them for distortion onto a $2\times$ oversampled grid, using the drizzle algorithm \citep{fru02} with a pixel reduction scale of 0.8. The PSF FWHM was $\sim80$ mas, limited by the 40 mas scale. Both of our fields have bright ($r < 18.1$) stars within $\sim16\arcsec$ of the galaxies that can be used for PSF determination, and our exposures were kept short enough to avoid saturating on these stars. Observations of a globular cluster field on the same night indicate that, for our purposes, PSF variations over this distance were negligible. Comparison of previous similar AO observations on a $z\sim2.5$ galaxy with {\it HST} NICMOS observations of the same galaxy gave essentially identical values for the \sersic\ $n$ index and $R_e$ \citep{sto08}, so we can have reasonable confidence in the detailed results from our AO imaging.

\section{Results}
\subsection{Spectral-Energy Distributions and Stellar Populations}
\begin{figure}[!bt]
\epsscale{1.0}
\plotone{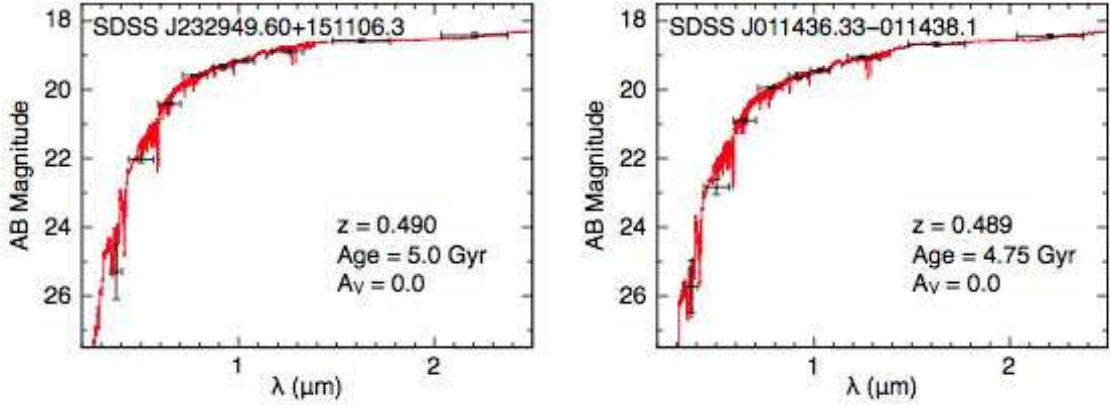}
\caption{Hyper-z best-fit SEDs, constrained by the spectroscopic redshifts for the galaxies and selecting among Charlot-Bruzual (2007; private communication) instantaneous burst models with 0.4, 1.0, and 2.5 solar metallicities and \citet{cha03} initial mass functions. \jtt\ and \jzo\ were fit best by 0.4 solar and 1.0 solar models, respectively.}\label{seds}
\end{figure}

We first show in Fig.~\ref{seds} the spectral-energy distributions (SEDs) for \jtt\ and \jzo, based on their SDSS/UKIDSS magnitudes. Because of the possibility that the zeropoints of the two surveys may differ at the few percent level \citep{hew06}, we have imposed a minimum uncertainty of 5\%\ on the magnitudes. We have used a modified version of the photometric redshift code {\sc Hyper-z} \citep{bol00} to fit SEDs to the magnitudes, where we have constrained the redshifts to be those we have determined from the spectroscopy described below. We explored a range of models, including both instantaneous-burst models and models with exponentially decreasing star-formation rates, all with a range of \citet{cal00} reddening and metallicities of 0.4, 1.0, and 2.5 solar. For both galaxies, the best fitting SED was from an instantaneous-burst model with no reddening. For \jtt, the model had an age of 5.0 Gyr and a metallicity of 0.4 solar, while for \jzo, the model had an age of 4.75 Gyr and solar metallicity. The indicated photometric masses, determined from the current mass in stars from the {\sc Hyper-z} model fits, are $3.9\times10^{11} M_{\odot}$ for \jtt, and $3.4\times10^{11} M_{\odot}$ for \jzo.

\begin{figure}[!bt]
\epsscale{1.0}
\plotone{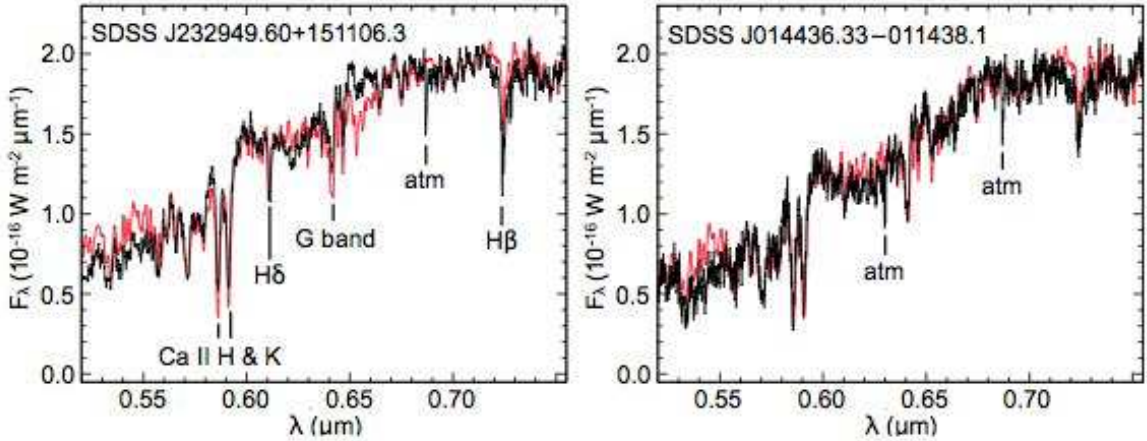}
\caption{Spectra of \jtt\ and \jzo\ (black traces) superposed on the spectral synthesis models that best fit the SDSS/UKIDSS photometry (red [gray] traces), as shown in Fig.~\ref{seds}.}\label{spec}
\end{figure}

Figure~\ref{spec} shows the LRIS spectra of \jtt\ and \jzo, compared with the models that give the best fits to the photometry, as shown in Fig.~\ref{seds}. Here we have simply scaled the models to the spectra (to account for slit losses). The redshifts for these two galaxies are fortuitously virtually the same, with \jtt\ having $z=0.490$ and \jzo\ having $z=0.489$.

\subsection{Morphologies}\label{morphsec}
\begin{figure}[!tb]
\epsscale{0.8}
\plotone{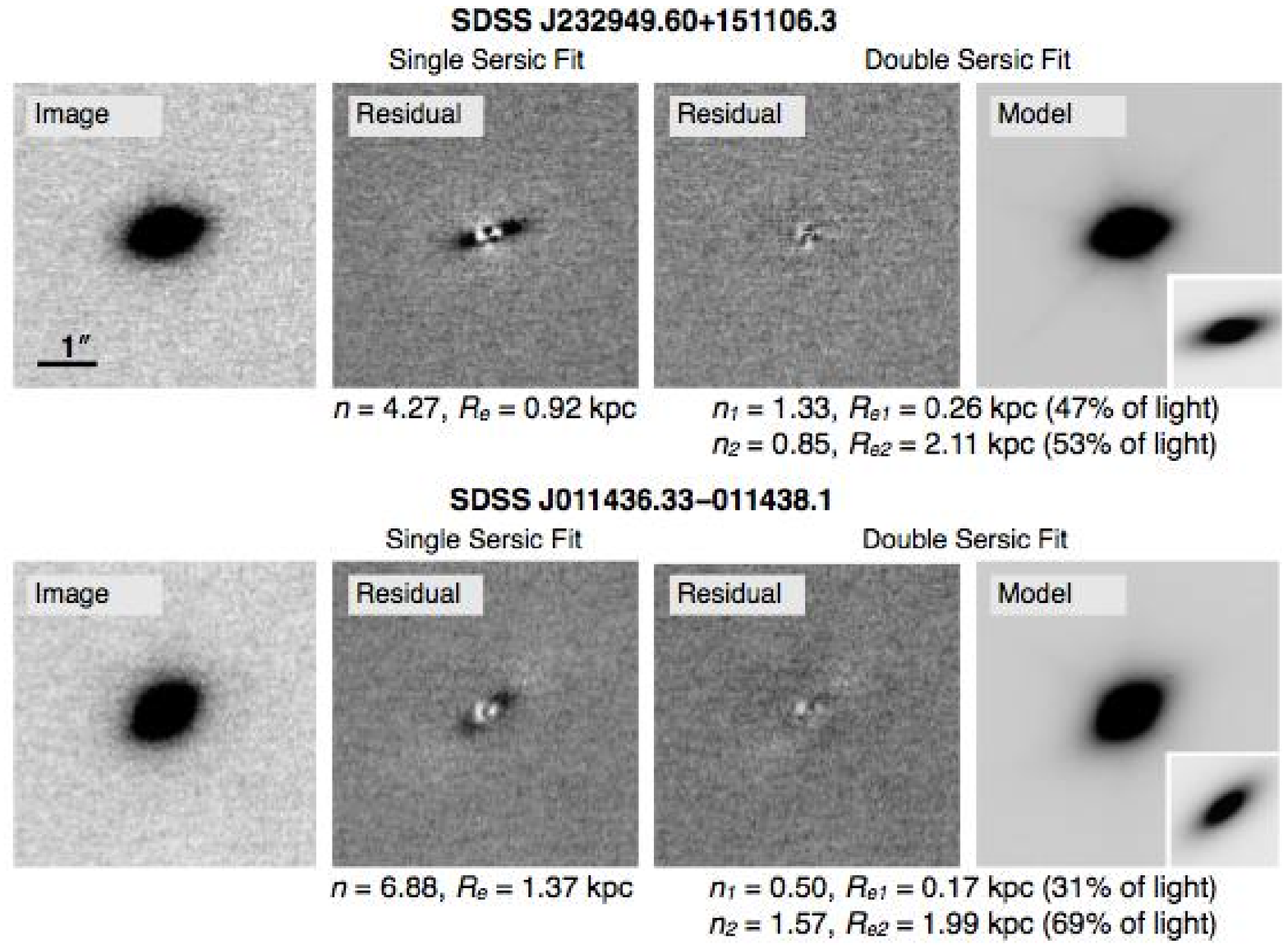}
\caption{Keck LGSAO images and model fits for \jtt\ and \jzo. For each galaxy, the panels from left to right show the original LGSAO image, the residual from the best-fit single-\sersic\ model, the residual from the best-fit double \sersic\ model, and the double \sersic\ model itself, at the same contrast as the original image. The insets in the model panels show the models without convolution with the PSFs, at slightly lower contrast. The \sersic\ index $n$ and the major axis $R_e$ are given below the corresponding model panels.}\label{morph}
\end{figure}

From our LGSAO imaging, we were able to explore the morphologies and radial-surface-brightness profiles of \jtt\ and \jzo. Using the model-fitting application {\sc galfit} \citep{pen02} and PSFs from nearby stars in the field, we first fit single \sersic\ models to the galaxy images, as shown in Fig.~\ref{morph} (second panels). These fits give \sersic\ $R_e = 0.92$ kpc for \jtt\ and $R_e = 1.37$ kpc for \jzo, in the range of the compact passive galaxies found at high redshifts, although not as small as the most extreme examples, which have $R_e \sim 0.5$ kpc. These single-component fits also leave systematic residuals with peak values $\sim2$\% of the peaks of the original images. We therefore explored 2-component fits, as shown in Fig.~\ref{morph} (third and fourth panels). These, as expected, show much smaller residuals; the main point, however, is that the residuals show little or no systematic structure.

An interesting feature of these fits is that single \sersic\ models give \sersic\ indices $n>4$, typical of spheroidal galaxies. At high redshifts, the small but systematic residuals left by these fits would fade into the noise. For the two-component \sersic\ fits, which give much more satisfactory residuals, the \sersic\ indices for both components are closer to exponential profiles. Furthermore, for the more extended components of our 2-component \sersic\ fits, the projected axial ratios are fairly small: $b/a = 0.34$ for SDSSJ2329 and 0.40 for SDSSJ0114. We have tried replacing the extremely compact central components with both a PSF component and a small $r^{1/4}$-law bulge, but the fits were significantly worse for both of these models.

\begin{figure}[!t]
\epsscale{1.0}
\plottwo{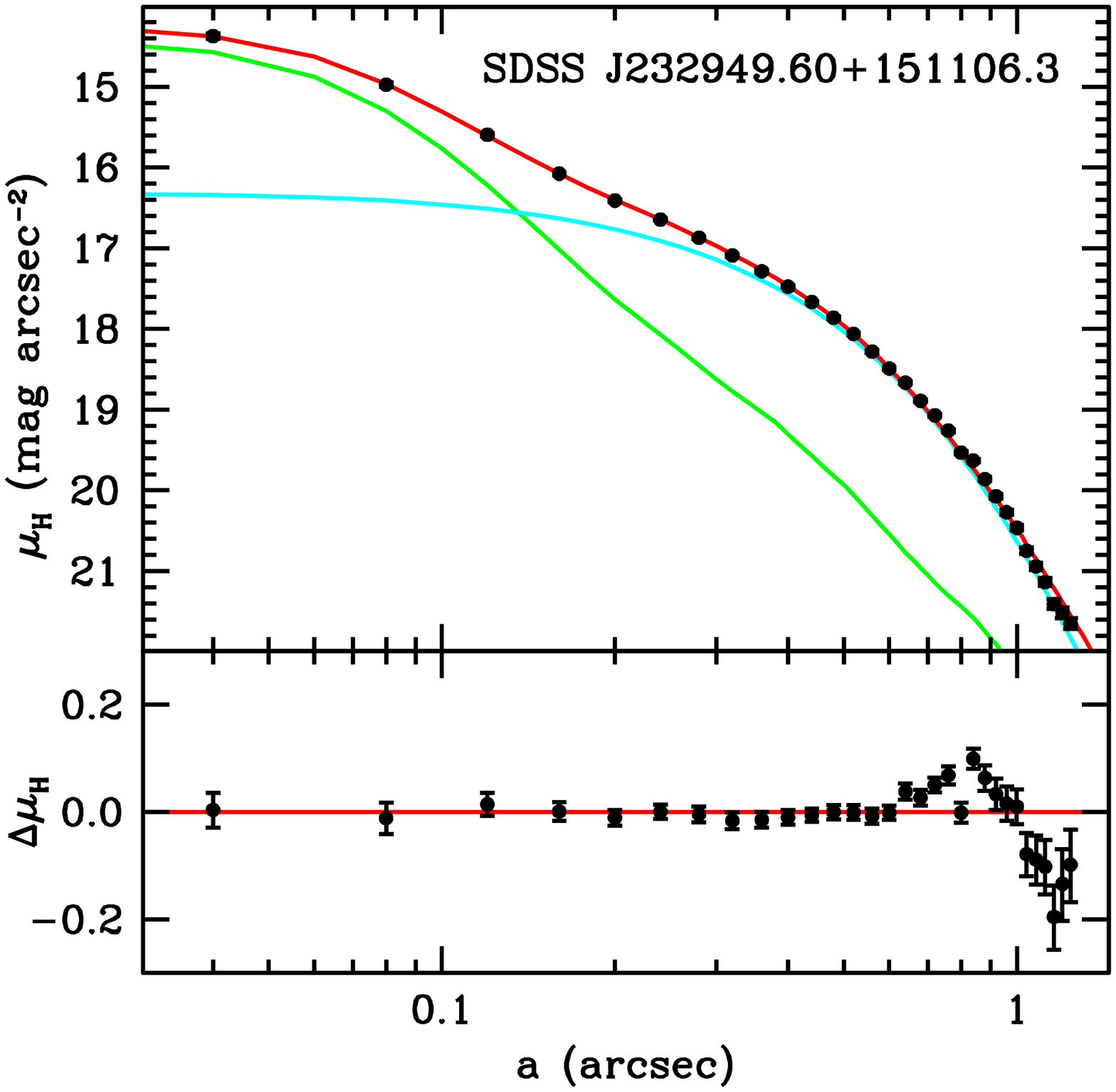}{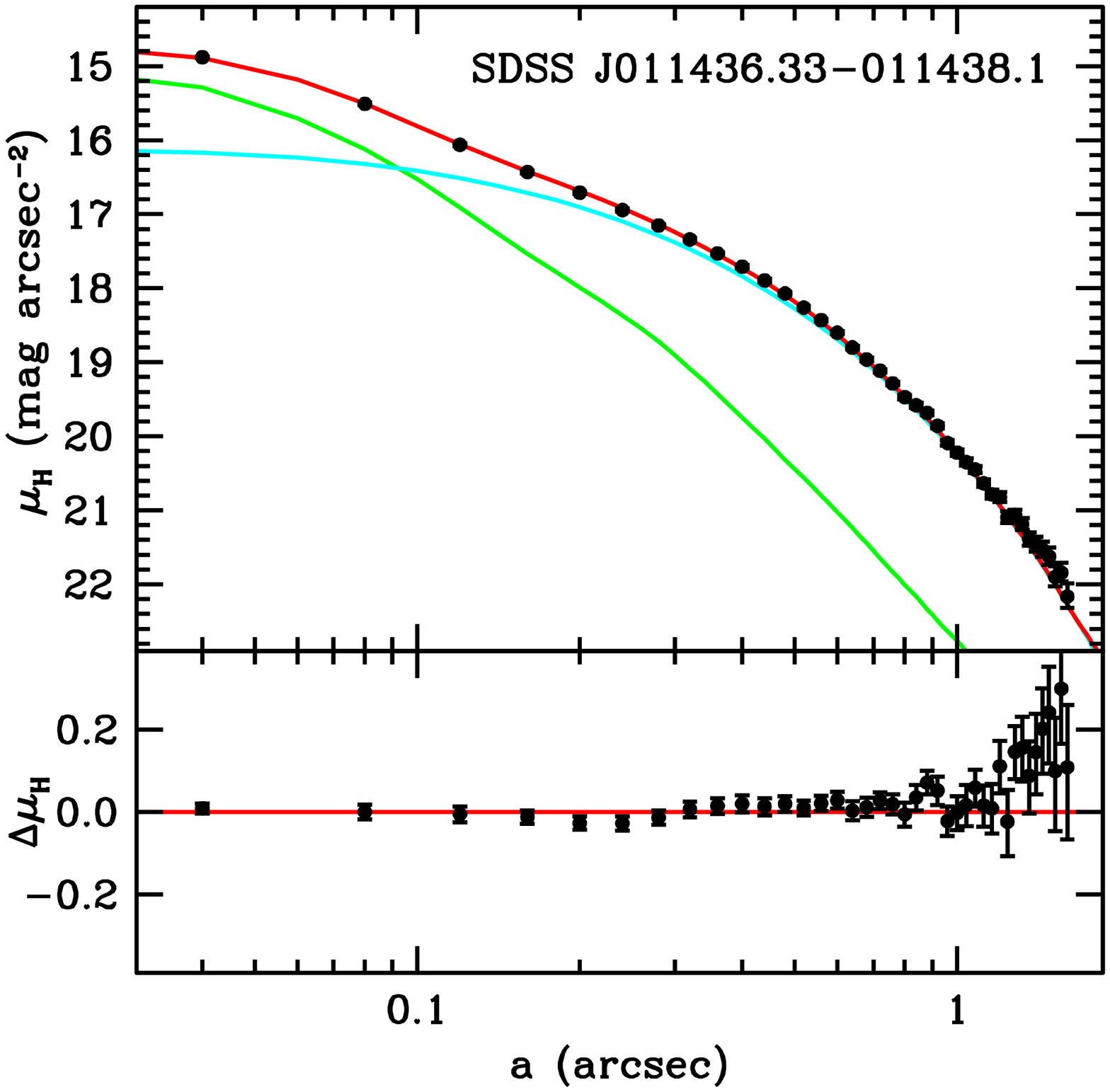}
\caption{Radial-surface-brightness profiles for \jtt\ and \jzo, from the LGSAO images shown in Fig.~3. The solid red (gray) lines show the best \sersic\ 2-component fit, with the green and blue (light gray) curves showing the individual components. The lower panels show the differences between the annulus photometry and the model fits.}\label{radsb}
\end{figure}

\begin{figure}[!ht]
\epsscale{0.5}
\plotone{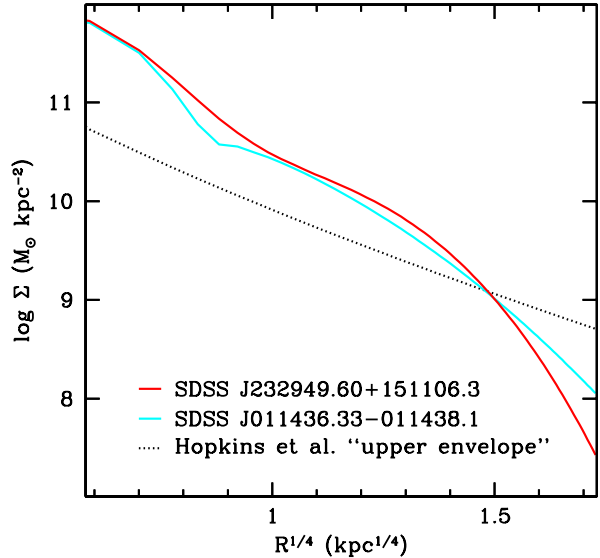}
\caption{Mass-surface-density profiles for \jtt\ and \jzo, assuming that the photometric masses derived from the spectral-synthesis model fits are correct. The radial coordinate $R$ is along the semi-major axis, and the surface densities are measured in elliptical annuli from the 2-component \sersic\ models, without convolution with the PSFs, which should closely approximate the true global morphologies of the galaxies beyond a radius of $\sim300$ pc (corresponding to $R^{1/4}\sim0.75$ kpc$^{1/4}$). The dotted line shows the ``upper-envelope'' profile of \citet{hop09}, which gives the approximate upper limit to the mass density at a given physical radius from a sample of massive elliptical galaxies at low redshifts. The range shown is $0.11 < R < 8.8$ kpc.}\label{sdprf}
\end{figure}

Figure~\ref{radsb} shows the radial-surface-brightness profiles along the major axes, derived from elliptical annuli sampling of the images, along with the best 2-component model fits. Figure~\ref{sdprf} shows the corresponding mass surface-density profiles (assuming a \citealt{cha03} IMF), together with the ``upper-envelope'' profile of \citet{hop09}, which represents an approximate upper limit to the mass surface density at each radius for any galaxy from an ensemble of massive elliptical galaxies in the local universe.

\subsection{Velocity Dispersions}\label{veldisp}

We have estimated the velocity dispersions from our spectra via the direct method, using spectra of G giant stars from \citet{val04} as templates. After re-binning all of the spectra to velocity units, we broadened the template spectra with an appropriate Gaussian to bring them to the same resolution as our de-redshifted spectra. For each galaxy, a grid of template spectra was prepared with a range of Gaussian broadening parameters, and the scaled template spectra were subtracted from the galaxy spectra. After subtraction of a low-order spline to take care of any local mismatches between the template and the galaxy spectra, the $\chi^2$ statistic was calculated for each broadening parameter. The minimum $\chi^2$ value was found by fitting a curve to $\chi^2$ values in the vicinity of the minimum.
In order to estimate the errors, we did Monte-Carlo simulations, creating 100 mock spectra that simulated both the random and correlated noise properties of the actual spectra for each galaxy. 

This procedure gives $\sigma = 291 \pm 16$ \kms\ for \jtt\ and $\sigma = 254 \pm 21$ \kms\ for \jzo. For spheroids, velocity dispersions can be translated to estimates of the dynamical mass by $M_{dyn} = kR_e\sigma^2/G$, where $k$ is a parameter that depends on the properties of the stellar velocity distribution function for the galaxy, as well as other factors, such as the region of the galaxy covered by the spectrum. It is often taken to be $\sim5$ \citep[\eg][]{ber09}. With this assumption, we find that $M_{dyn} = 0.91\times10^{11} M_{\odot}$ for \jtt\ and $1.0\times10^{11} M_{\odot}$ for \jzo, where we have used the single-\sersic\ estimates for $R_e$. These values are about 30\%\ of the masses estimated from the stellar population fits. The discrepancy could be attributed to problems with the assumption of a standard IMF or concerns about the value used for $R_e$ (although, since the latter enters linearly, it is unlikely to make up the entire difference).

However, it is also possible that it is simply wrong to assume that a procedure normally applied to ellipticals is wholly relevant in these cases. Both the \sersic\ indices of the best-fitting models and the axial ratios of more extended components fall outside the range of normal spheroids, so our ``velocity dispersion'' could actually be dominated by bulk velocities of one kind or another. We have modeled rotation curves, under the assumption that both components of each of these galaxies are rotating disks, that the mass-to-light ratio is constant, and that the disks can be approximated by flattened spheroids. This last assumption simplifies the calculation and is a close enough approximation to the exponential disk case to show the general form of the expected profile fairly accurately \citep[see, \eg][]{bin87}. These models show that even for total masses $<0.1$ of the photometric mass we have found, we would expect to see double lines at the resolution of our integrated spectra, which are clearly not present. Another possibility is that the galaxies are prolate, with mostly radial orbits aligned with the long axis. Further work, including resolved spectroscopy, is needed to explore this issue.

\section{Discussion}
The two galaxies we discuss here show many of the characteristics of the luminous compact passive galaxies found at high redshift, and their relative proximity allows us to explore them in much greater detail than is possible for the high-redshift cases. However, they cannot actually be survivors of the population found at $z>2$, since their stellar population ages indicate that most of the stars in them were formed at $z\sim1.8$. Nevertheless, it is still quite possible that they have had similar formation histories to those of the high-redshift population, although perhaps in less dense environments, where formation processes might have been delayed. Assuming that we can take \jtt\ and \jzo\ as mirroring the properties of the compact high-redshift galaxies, what implications do they suggest?

\begin{itemize}{
\item Our surface-brightness profiles shown in Fig.~\ref{radsb} cover a range of over 6 magnitudes over a semi-major axis of $\sim6$ kpc. The fact that we can probe these galaxies to much lower surface-brightness levels than we can their high-redshift counterparts shows that, in at least these cases, their compactness is unlikely to be explained by suggestions that we are simply missing low-surface-brightness outer regions \citep[\eg][]{hop09}.

\item The necessity of two-component fits to avoid systematic residuals indicates that these galaxies have already undergone considerable dynamical evolution. They are not as simple systems as might have been assumed from the available data on their high-redshift counterparts.

\item The best fitting models are superpositions of components with surface-brightness profiles close to exponentials, even though the single-component models have profiles close to $r^{1/4}$ laws. In addition, the more extended components of the reconstructed models have high axial ratios, as seen in the insets to Fig.~\ref{morph}. These conclusions line up with evidence that luminous passive galaxies at high redshifts often have exponential profiles and an apparent disk-like appearance \citep{sto04,sto07,sto08}. However, the lack kinematic indications of a disk in our integrated spectra suggest that these galaxies may not, in fact, be rotationally supported. Prolate morphologies with strongly anisotropic stellar velocity fields remain a possibility.

\item The two-component models that best fit the luminosity profiles indicate that slightly more than half of the light from \jtt\ and \jzo\ comes from structures that have $R_e \sim 2$ kpc, still small, but considerably larger than the $R_e$ found from single-component models. On the other hand, $\gtrsim1/3$ of the light comes from extremely compact structures with $R_e\lesssim250$ pc. Assuming a standard \citet{cha03} initial mass function (IMF), our spectral synthesis model fits in Fig.~\ref{seds} imply masses of $\gtrsim10^{11} M_{\odot}$ for these extremely compact cores.
}
\end{itemize}

\citet{bez09} and \citet{hop09} present plausible and convincing arguments that the luminous passive compact galaxies at high redshifts are locally present as the core regions of the most massive present-day elliptical galaxies, which have presumably acquired their extended envelopes through mostly dry mergers with more recently formed and fluffier galaxies. \citet{hop09} define an ``upper envelope'' to the surface mass density profiles of massive local ellipticals and show that the apparent mass surface density of the high-redshift compact galaxies rarely rises above this profile at the same physical radii. We have plotted this upper-envelope curve in Fig.~\ref{sdprf}, for comparison with the mass-surface-density curves for our best-fit photometric mass models for SDSSJ2329 and SDSSJ0114. The surface densities for both models are a factor of $\sim3$ higher than the \citet{hop09} upper envelope even out to $R=2$ kpc. This factor rises to $\sim10$ in the inner regions. If this difference can be trusted (and if it is representative of the high-redshift counterparts to these galaxies), it indicates that there still needs to be some amount of ``puffing up'' of these objects to bring their central surface densities into line with those of massive ellipticals at low redshift. Of course, our mass-surface-density profiles might be overestimated if our total mass estimates are too large because (1) we have grossly overestimated the ages of the stellar populations (we believe this to be unlikely) or (2) the IMFs for the stellar populations are more top-heavy than the \citet{cha03} IMFs we have assumed. If the actual masses of these galaxies were determined to be more in line with our velocity dispersion measurements, this latter possibility would provide the most likely explanation.  Future observations of these and other similar galaxies at moderate redshifts should give us insight into these issues.

\acknowledgments
We thank the anonymous referee for a careful and thoughtful reading of the original version of this {\it Letter} and for offering numerous suggestions to improve both its substance and presentation. We also thank Phil Hopkins for his comments on the earlier version.
The UKIDSS project is defined in \citet{law07}. UKIDSS uses the UKIRT Wide Field Camera (WFCAM; \citealt{cas07}) and a photometric system described in \citet{hew06}. The pipeline processing and science archive are described in \citet{ham08}. Funding for the SDSS has been provided by the Alfred P. Sloan Foundation, the Participating Institutions, the National Science Foundation, the U.S. Department of Energy, the National Aeronautics and Space Administration, the Japanese Monbukagakusho, the Max Planck Society, and the Higher Education Funding Council for England. The SDSS Web Site is http://www.sdss.org/.

{\it Facilities:} \facility{Keck:I (LRIS)}, \facility{Keck:II (LGSAO/NIRC2)}, \facility{Sloan (SDSS)}, \facility{UKIRT (UKIDSS)}




\end{document}